\documentclass[epj]{webofc}
\usepackage[varg]{txfonts}   % Web of Conferences font
\wocname{EPJ Web of Conferences}
\woctitle{INPC 2013}
\begin{document}
\title{Nuclear stopping for heavy-ion induced reactions in the Fermi Energy range : from 1-Body to 2-Body dissipation}
%
% subtitle is optional

\author{Olivier Lopez\inst{1}\fnsep\thanks{\email{lopezo@in2p3.fr}} \\(INDRA collaboration)}

\institute{Laboratoire de Physique Corpusculaire de Caen, ENSICAEN, Université de Caen, CNRS-IN2P3, 
Boulevard Maréchal Juin, 14050 Caen, France}

\abstract{We address the stopping in heavy-ion induced reactions around the Fermi energy using central collisions 
recorded with \emph{INDRA} $4\pi$ array. The stopping is minimal around the Fermi energy and corresponds to the 
disappearance of the Mean-Field effects and the appearance of nucleon-nucleon elastic collisions. This phenomenon is 
attributed to a change in the dissipation regime going from 1-body (Mean-Field) dissipation at low incident energy to 2-
body (nucleonic collisions) above the Fermi energy. A connection to the in-medium transport properties of nuclear 
matter is proposed and comprehensive values of the nucleon mean free path $\lambda_{NN}$ and the nucleon-nucleon cross 
section are extracted.
}
\maketitle
\section{Introduction}
\label{intro}
The study of transport phenomena is of utmost importance for understanding the fundamental properties of nuclear 
matter\cite{Durand2006}. They are also critical in the description of the supernova collapse and the formation of a 
neutron star\cite{Lattimer2004}. Transport properties of nuclear matter are also one of the basic ingredients for 
microscopic models\cite{Andronic2006}. In this study, we are looking at the global energy dissipation achieved in heavy-
ion induced reactions in the Fermi energy domain. We are using the large dataset available in this energy range for 
symmetric systems recorded with the $4\pi$ array \emph{INDRA}\cite{Pouthas1995}. We are looking at central collisions, \
emph{i.e.} corresponding to the maximal overlap between the 2 incoming nuclei. Doing so, we can extract information 
concerning the stopping encountered in such collisions and relate it tentatively to the nucleon mean free path in the 
nuclear medium as well as the in-medium quenching factor for the nucleon-nucleon cross section. 

\section{Stopping for central collisions}
\label{sec-1}
We are in this work focused on the \emph{INDRA} data coming from central collisions, selected through the total 
multiplicity of charged particle as done in \cite{Lehaut2010}. The data span a large body of symmetric systems, with 
total mass between 70 and 500, and incident energies covering the full range of the Fermi domain $E_{inc}/A=10-100~MeV$.
 To evaluate the degree of stopping achieved in central collisions, we are using the energy isotropy ratio $R_E$ 
defined as : 

\begin{equation}
 R_E=\frac{1}{2} \frac{\sum_i E_i^{\perp}}{\sum_i E_i^{//}}
\end{equation}

\begin{figure}[h]
\centering
\sidecaption
\includegraphics[width=7cm,clip]{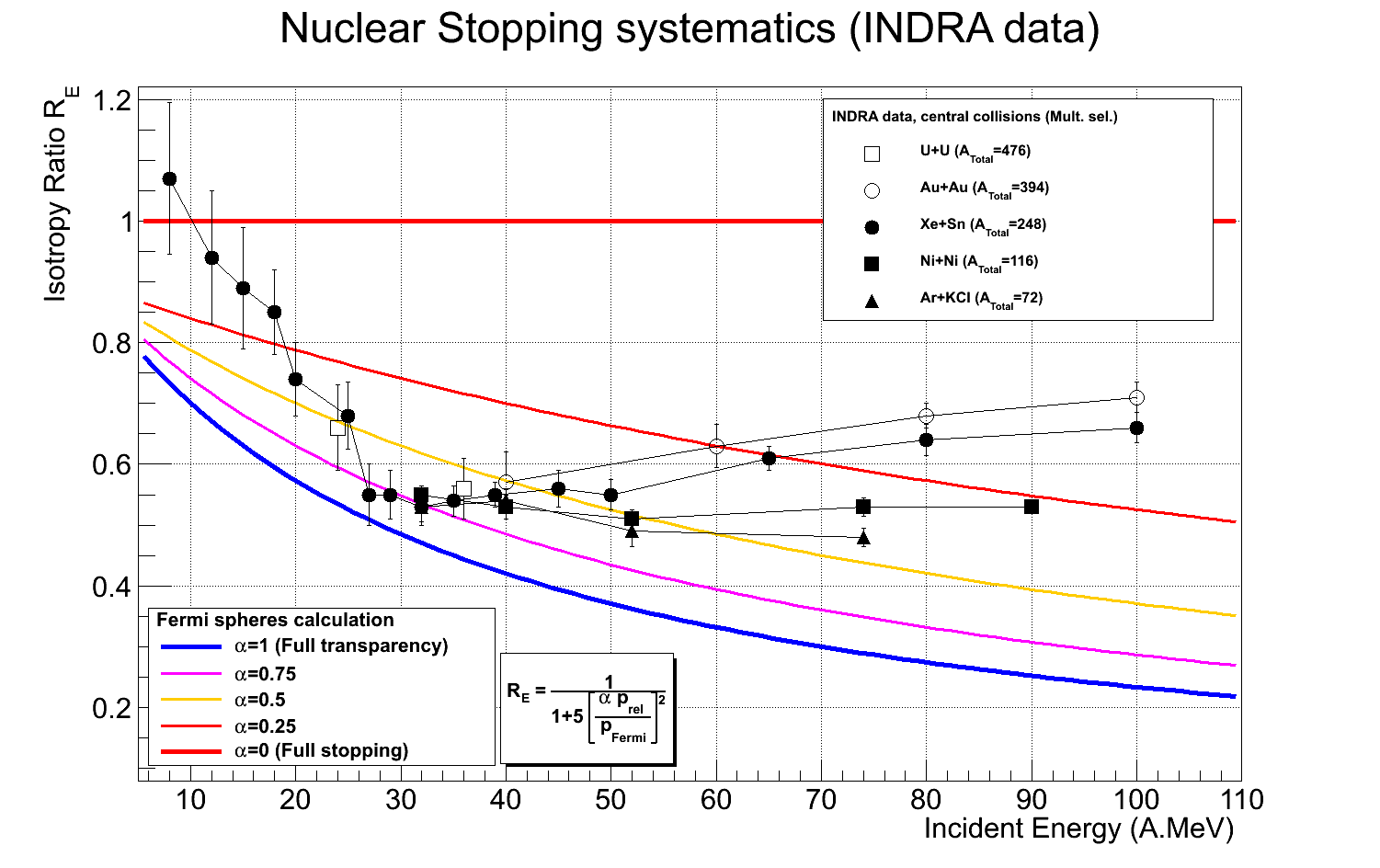}
\caption{Isotropy ratio $R_E$ as a function of incident energy for central collisions in symmetric systems.}
\label{stopping1}       
\end{figure}

The sum $i$ runs over the total number of charged particle in the event, $E_i{\perp}=E_i \sin^2(\theta_i)$  and $E_i{\
perp}=E_i \cos^2(\theta_i)$ are respectively the transverse and parallel energies in the center-of-mass of the reaction,
 $E_i$ and $\theta_i$ being the total $CM$ energy and polar angle. By construction, this ratio is taking values between 
$0$ and $1$, $1$ corresponding to an isotropic emission. We present in figure \ref{stopping1} the mean isotropy ratio 
$<R_E>$ calculated for all systems as a function of the incident energy. 

\begin{figure}[h]
\centering
\sidecaption
\includegraphics[width=7cm,clip]{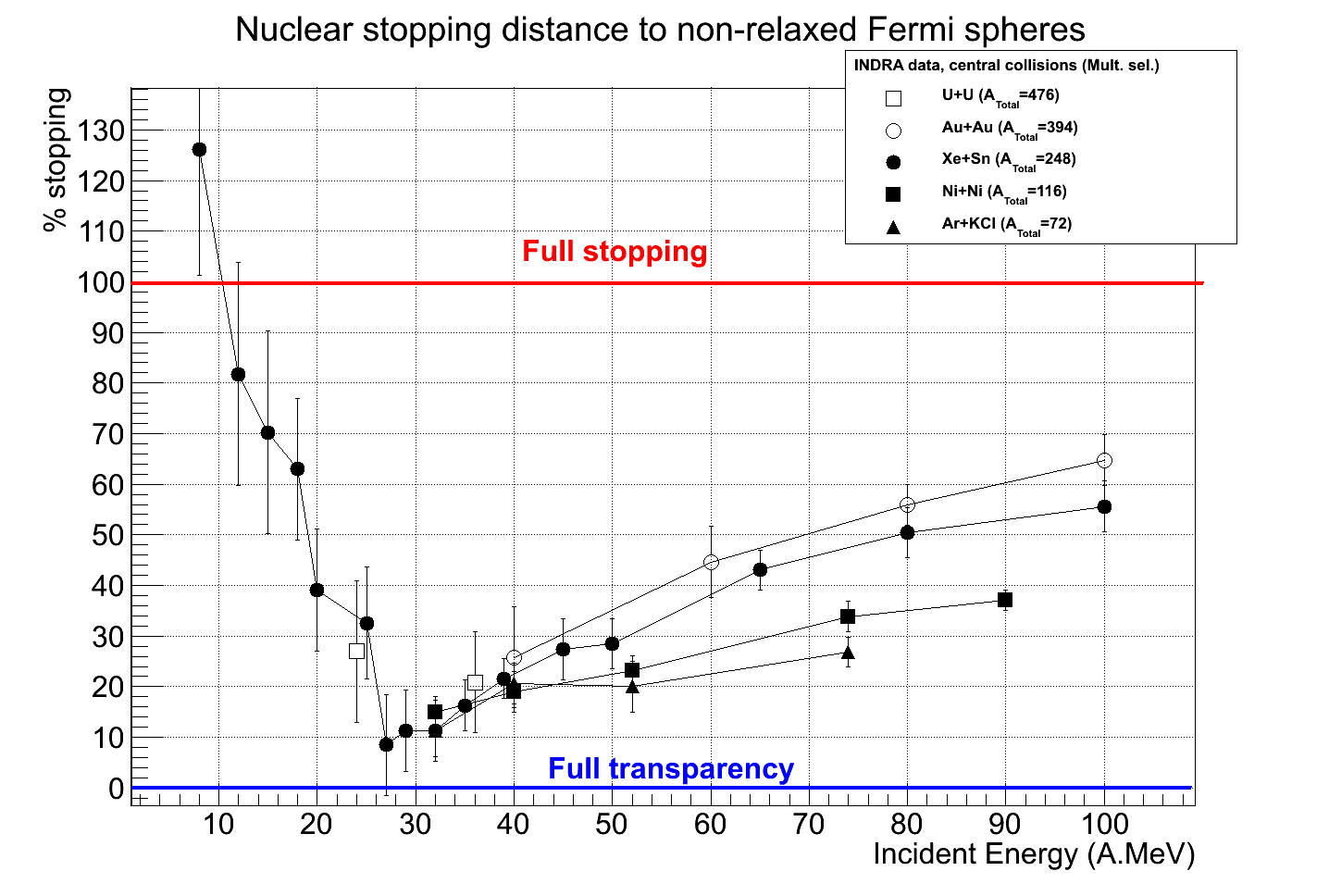}
\caption{Normalized distance $d$ as a function of incident energy.}
\label{stopping2}     
\end{figure}

The coloured curves on figure \ref{stopping1} correspond to several analytical calculations 
associated to the entrance channel; we consider here 2 Fermi spheres separated by the relative 
momentum $\alpha P_{rel}$ corresponding to the relative energy of the entrance channel. A quite straightforward 
calculation shows that the energy isotropy ratio $R_E^{EC}(\alpha)$ is defined as : 

\begin{equation}
 R_E^{EC(\alpha)}=\frac{1}{1+5(\frac{\alpha P_{rel}}{P_{Fermi}})^2}
 \end{equation}

 where $P_{Fermi}$ is the Fermi momentum at saturation density ($P_{Fermi}=266~MeV/c$), $P_{rel}=\mu V_{rel}$ is the 
relative momentum, $\mu$ is the reduced mass and $V_{rel}$ the relative velocity. $\alpha$ is a free parameter which 
represents the degree of dissipation observed in the collision; $\alpha=0$ means full dissipation while $\alpha=1$ 
corresponds to the sudden approximation with no dissipation. The data are then dispatched between these two extreme 
values as shown in Fig. \ref{stopping1}. We note that the closest distance to the \emph{no dissipation} scenario ($\
alpha=1$) corresponds to an incident energy slightly below the Fermi energy, here around $30A~MeV$. To better quantify 
this effect, we compute the distance $d=\frac{R_E-R_E^{EC}(\alpha=1}{R_E^{EC}(\alpha=0)-R_E^{EC}(\alpha=1)}$ between 
the data and the entrance channel, normalized to the distance between the values corresponding to $\alpha=0$ and $\
alpha=1$. The results are displayed in figure \ref{stopping2}. A minimum around $E_{inc}/A=30~MeV$ is clearly visible, 
altogether with the decrease from barrier energy to Fermi energy and the increase above the Fermi energy. This is 
attributed to the change in the energy dissipation regime as already explained in \cite{Lehaut2010}. 
  
\begin{figure}[h]
\centering
\sidecaption
\includegraphics[width=7cm,clip]{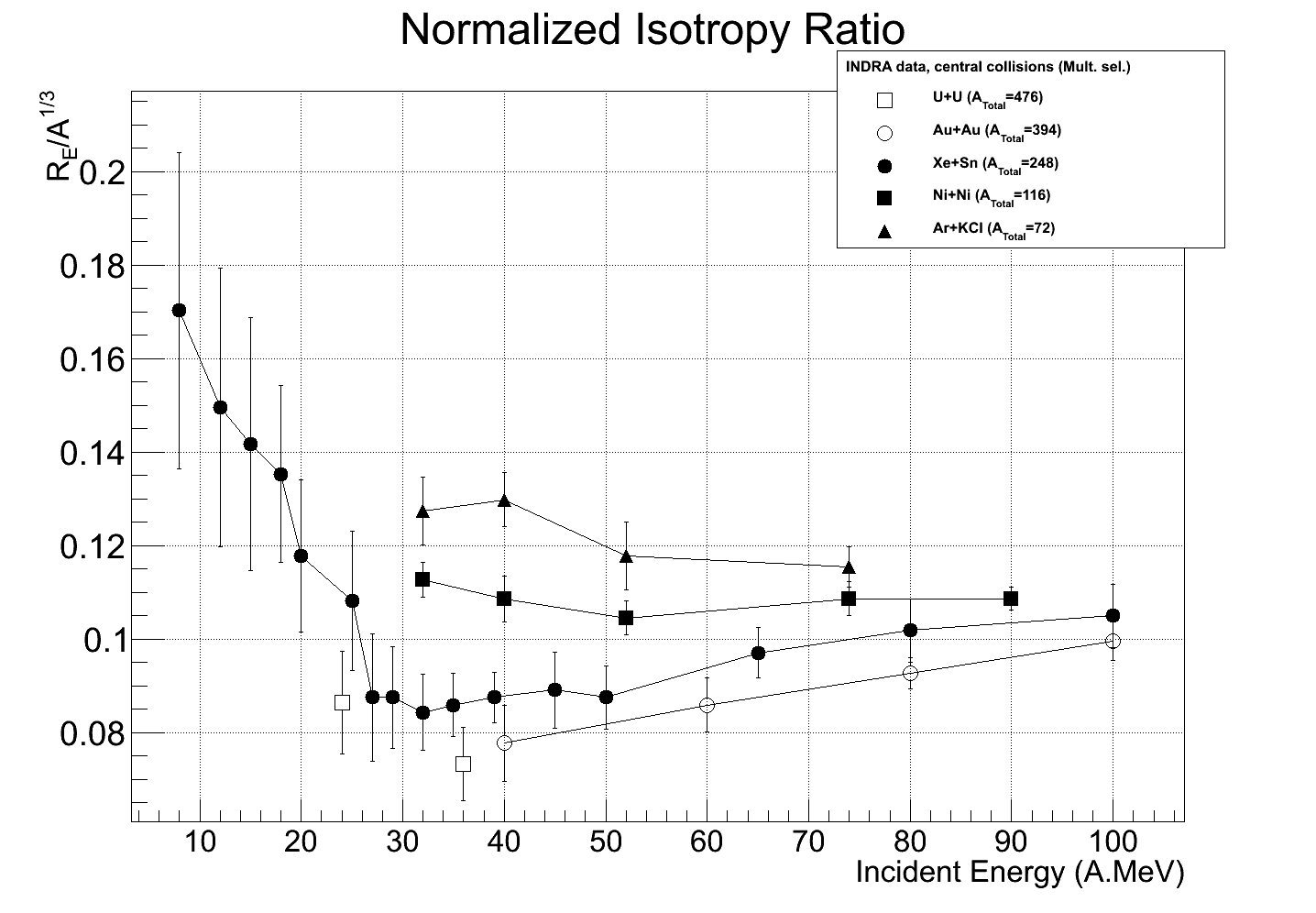}
\caption{Reduced isotropy $R_E'=R_E/L$ as a function of incident energy. See text for details. }
\label{stopping3}     
\end{figure}

Above the Fermi energy, in the nucleonic regime, we also observe a nice mass hierarchy concerning the stopping: the 
heavier the system, the larger the stopping is. This is fully consistent with a \emph{Glauber} picture where the degree 
of stopping is closely related to the number of participants. To go further, we can normalize the isotropy ratio to the 
characteristic size of the system, here the radius $L=r_0 A^{1/3}$ with $r_0=1.2~fm$. Figure \ref{stopping3} displays 
the results for this reduced quantity.

We observe that all systems seem to collapse to the same value at high energy, well above the Fermi energy, above 
$100A~MeV$. This is here an indication that the stopping is related to the characteristic size $L$ of the system, and 
is then connected to the nucleon mean free path $\lambda_{NN}$ in the nuclear medium; to be more specific, we have 
performed simple Monte Carlo 
simulations of 2 Fermi spheres using the \emph{Glauber} approach to check this assumption. We have then obtained the 
following relationship between the in-medium mean free path $\lambda_{NN}$ and the normalized distance $d$ : 

\begin{equation}
\lambda_{NN} \approx \frac{L}{d^{2/3}}
\end{equation}

\section{Nucleon Mean Free Path in the medium}

\begin{figure}[h]
\centering
\sidecaption
\includegraphics[width=7cm,clip]{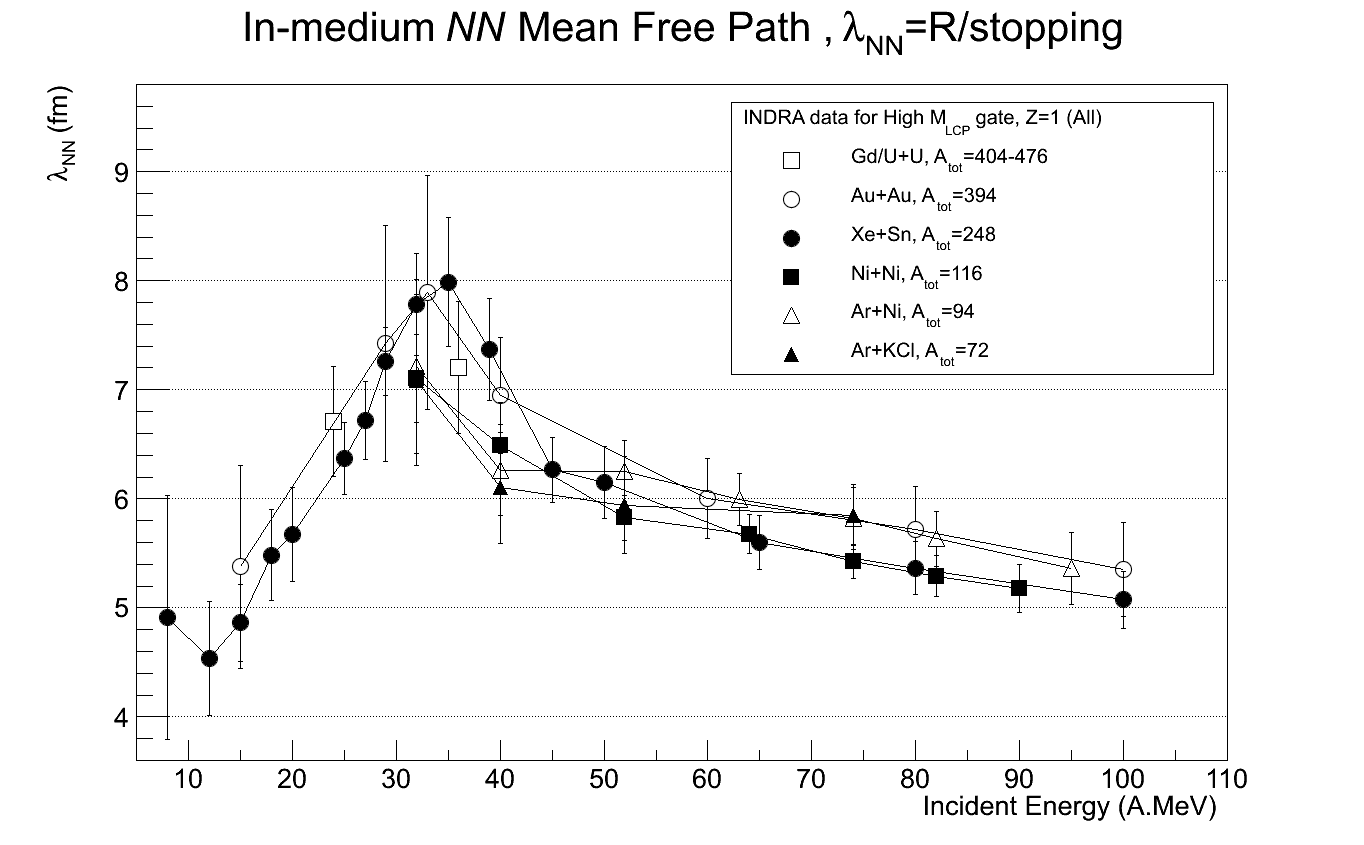}
\caption{Nucleon Mean Free Path $\lambda_{NN}$ as a function of incident energy.}
\label{stopping4}     
\end{figure}

Applying the findings of the previous section concerning $\lambda_{NN}$, we obtain the results displayed by Fig. 
\ref{stopping4}.  It is interesting to note that all curves exhibit the same behaviour whatever the total mass. We observe a maximum at incident energy corresponding to minimal stopping, here at $E_{inc}=30A~MeV$. The decrease at low incident energy (below $30A~MeV$), can be attributed to the fact that the Mean-Field dissipation is well active in this energy domain, and consequently the normalized distance from the entrance channel is not anymore correct; in this domain, we should adjust the $\alpha$ parameter to describe at best the energy dissipation. In the limited scope of the present 
study, we are mainly interested on the nucleonic regime. Going back to this incident energy domain, we observe a 
decrease from $\lambda_{NN}\approx 8~fm$ to  $\lambda_{NN}=5~fm$. It is nice to see that the value at high energy is in 
agreement with recent \emph{Dirac-Brueckner-Hartree-Fock} calculations using realistic interactions \cite{Rios2012}.

\section{In-medium effects for nucleon-nucleon cross section}

Having the mean free path, we can now estimate the nucleon-nucleon cross section by taking the usual formula from the 
kinetic theory of gases : $\sigma_{NN}=1/(\rho \lambda_{NN})$ where $\rho$ is the nuclear density, taken as the 
saturation density $\rho=0.16~fm^{-3}$. To get the \emph{true} nucleon-nucleon cross section, we have to correct from 
the \emph{Pauli} blocking as a trivial (but essential) in-medium effect. To do so, we use the prescription of ref. \
cite{Kikuchi1968}. More sophisticated approaches concerning the \emph{Pauli} blocking estimation can be found in the 
litterature \cite{Chen2013}, but we consider here that the estimated correction factor is correct. We then construct 
the quenching factor $F=\sigma_{NN}^{in-medium}/\sigma_{NN}^{free}$ by taking the energy-dependent free nucleon-nucleon 
cross section $\sigma_{NN}^{free}$ available in the litterature \cite{Metropolis1958}.

\begin{figure}[h]
\centering
\sidecaption
\includegraphics[width=7cm,clip]{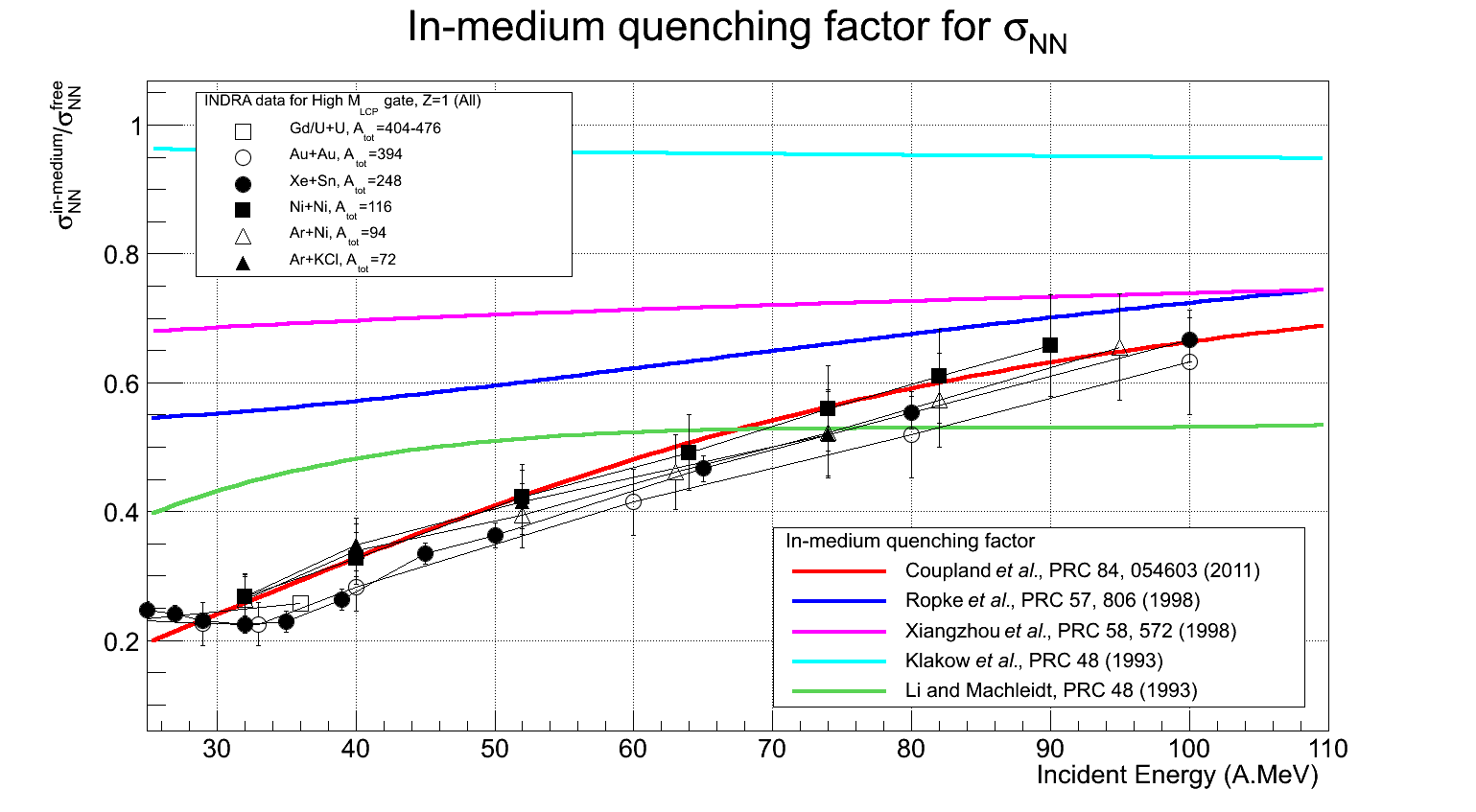}
\caption{In-medium correction factor $F$ for nucleon-nucleon cross section as a function of incident energy.}
\label{stopping5}     
\end{figure}

The results are visible on Fig. \ref{stopping5} for the quenching factor $F$. We display results only for the energy 
range above $E_{inc}/A=30~MeV$. The 'experimental' values are compared to standard theoretical prescriptions derived 
from different works and currently used in microscopic transport models \cite{Klakow1993,LiMachleidt1993,Schnell1998,
Xiangzhou1998,Coupland2011}. From the experimental results, we see that the medium effects are quenching almost all 
nucleon-nucleon collisions at low energy with $F\approx 20\%$. We obtain values close to $F=80\%$ at the highest 
incident energy, here at $100A~MeV$. It is interesting to note that all parametrizations, except the one of ref. \cite{
LiMachleidt1993}, are converging toward this value, indicating that the in-medium effects are correctly taken into 
account around $80A-100A~MeV$ incident energy. For the low energy domain, the parametrizations give quite different 
results. The only describing correctly the experimental data is the one of ref. \cite{Coupland2011}. This parametrizatio
n is based upon a phenomenological assumption made by \emph{Danielewicz} concerning the quenching factor : 

\begin{equation}
F= \frac{\sigma_0}{\sigma_{NN}^{free}} \tanh(\sigma_{NN}^{free}/\sigma_0)
\end{equation}

where $\sigma_0$ is the minimal nucleon-nucleon cross section and is geometrically related to the inter-particle 
distance in the nuclear medium, such as : $\sigma_0=0.85/rho_0^{2/3}$. It is worthwhile to mention that $\sigma_0$ uses 
the same functional dependence $\rho^{2/3}$ as the one deduced from the \emph{Glauber}-type of Monte Carlo simulations. 

\section{Conclusions}

In this study, we have estimated the nucleon mean free path and the related nucleon-nucleon cross section in the 
nuclear medium from the degree of stopping achieved in central collisions in the Fermi domain between $30A$ and 
$100A~MeV$. The estimated values for the mean free path are comprised between $8fm$ at $E_{inc}/A=30~MeV$ and $5fm$ at 
$E_{inc}/A=100~MeV$. The relatively large value around the Fermi energy suggests that full thermalization cannot be 
reached in such central collisions. These values are moreover in agreement with recent theoretical findings using 
microscopic approaches. In-medium effects, \emph{Pauli} blocking and \emph{high-order correlations} in medium, have 
also been experimentally inferred and are found to be large in the Fermi energy range; this is a clear indication that 
these effects have to be properly taken into account in any microscopic transport model. Concerning the quenching 
factor in nuclear medium, the best agreement is obtained by using the prescription of \cite{Coupland2011}. These 
results have nevertheless to be confirmed in a forthcoming analysis. Anyway, this analysis shows that the stopping can 
be used as a powerful probe to study transport properties of nuclear matter.

\end{document}